\documentclass{mn2e}
\usepackage{epsfig}
\usepackage{times}
\begin{document}

\title[black hole binary jets] 
{No evidence for black hole spin powering of jets in X-ray binaries}
\author[Fender et al.]  
{R. P. Fender$^{1,2}$\thanks{email:r.fender@soton.ac.uk}, E. Gallo$^3$, D. Russell$^4$\\
$^1$ School of Physics and Astronomy, University of Southampton, Highfield, Southampton, SO17 1BJ, UK\\
$^2$ Laboratoire d'Astrophysique de Grenoble, UMR 5571, CNRS and Universit\'e Joseph Fourier, F-38041 Grenoble, France\\
$^3$ MIT Kavli Institute for Astrophysics and Space Research, 70 Vassar  
St., Cambridge, MA 02139, USA\\
$^4$ Astronomical Institute `Anton Pannekoek', University of Amsterdam, P.O. Box 94249, 1090 GE Amsterdam, the Netherlands
\\
}
\maketitle

\begin{abstract}
In this paper we take the reported measurements of black hole spin for
black hole X-ray binaries, and compare them against measurements of
jet power and speed across all accretion states in these systems. We
find no evidence for any correlation between the properties of the
jets and the reported spin measurements.  These constraints are
strongest in the hard X-ray state, which is associated with a
continuous powerful jet. We are led to conclude that one or more of
the following is correct: (i) the calculated jet power and speed
measurements are wrong, (ii) the reported spin measurements are wrong,
(iii) there is no strong dependence of the jet properties on black
hole spin. In addition to this lack of observational evidence for a
relation between black hole spin and jet properties in stellar mass
black holes, we highlight the fact that there appear to be at least
three different ways in which the jet power and/or radiative
efficiency from a black hole X-ray binary may vary, two of which are
certainly independent of spin because they occur in the same source on
relatively short timescales, and the third which does not correlate
with any reported measurements of black hole spin. We briefly discuss
how these findings may impact upon interpretations of populations of
active galactic nuclei in the context of black hole spin and merger
history.
\end{abstract}

\section{Introduction}

\begin{figure}
\centerline{\epsfig{file=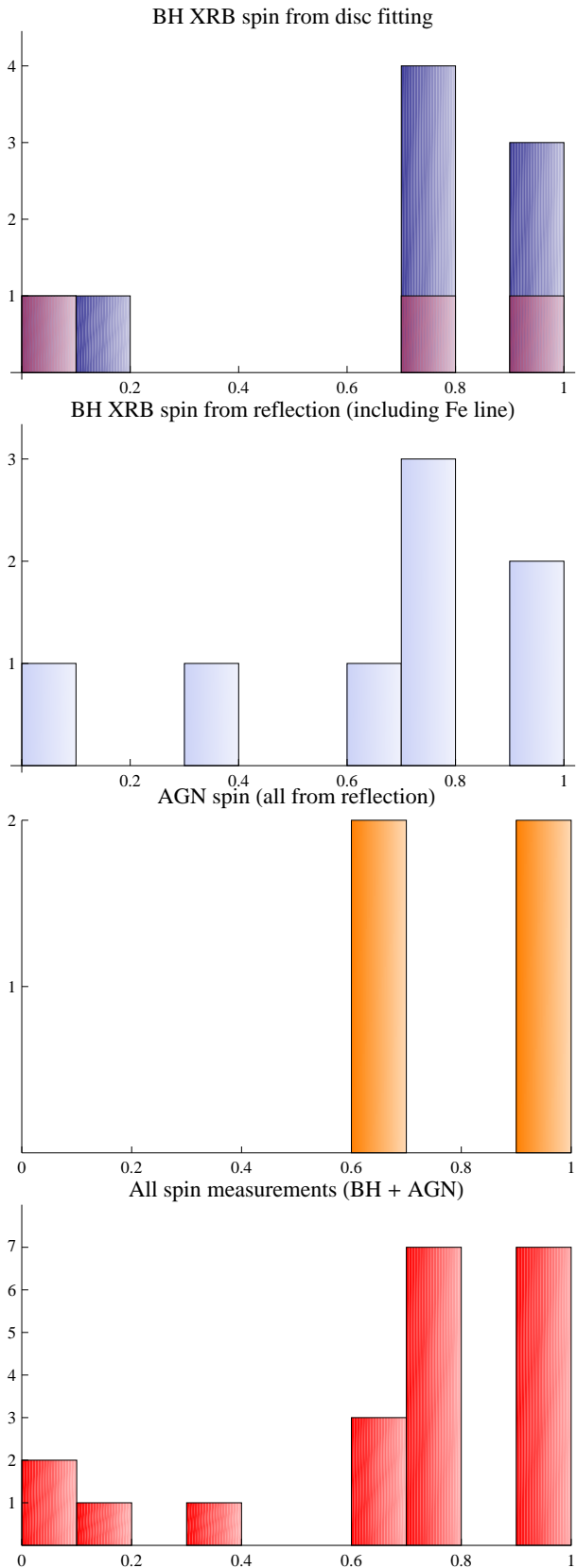, width=8cm, angle=0}}
\caption{A compilation of reported spin measurements for black
  holes. The first panel shows measurements for black hole X-ray
  binaries (BH XRBs) based on disc continuum fitting; the three purple
  measurements are the three different spins reported for the system
  GRS 1915+105. The second panel shows measurements for BH XRBs based
  on disc reflection (including iron line) fitting. The third panel
  shows measurements for AGN, all of which are based upon reflection
  fits. The fourth and final panel shows the sum total of the three
  previous histograms.}
\label{spincomp}
\end{figure}

Black holes remain one of the most bizarre and intriguing aspects of
astrophysics. In general relativity a black hole is entirely described
by only three parameters, mass, spin and charge. Mass is the easiest
of these parameters to measure (most accurately by observing the
orbits of other bodies around the black hole, such as a binary
companion, nearby stars, masers etc), and charge is generally supposed
to be unimportant, with astrophysical source electrically neutral on
average on macroscopic scales.  Black hole spin is not just a
curiosity; a spinning black hole has a smaller event horizon than a
non-rotating hole, and consequently a deeper gravitational well
outside of the horizon, potentially increasing the efficiency of
accretion. In addition, the rotational energy of spinning black holes
may be enormous ($\sim 30$\% $Mc^2$ for a maximally spinning black
hole), and could potentially be tapped as an energy source (Penrose
1969; Christodolou 1970). This concept was placed into the framework
of accretion by Blandford \& Znajek (1977) who investigated the
extraction of black hole spin by a magnetic field supported by an
accretion disc, and concluded that energy and angular momentum could
be extracted from the black hole in this way. This concept was
extended by MacDonald \& Thorne (1982), and more recently discussed by
Livio, Ogilvie \& Pringle (1999) who concluded that the likely
extraction of rotational energy of the black hole had been
overestimated. McKinney (2005) however arrived at the opposite
conclusion, deriving a very strong dependence of jet power on black
hole spin (see also De Villiers et al. 2005 for parallel work on
the influence of spin on jets from numerical simulations).  The most
frequent discussion of black hole spin is in the context of the
apparent radio loud:radio quiet `dichotomy' in active galactic nuclei
(e.g. Sramek \& Weedman 1980; Stocke et al. 1992; Miller, Rawlings \&
Saunders 1993; Xu, Livio \& Baum 1999), which may have an origin in
the powering of AGN jets by black hole spin (e.g. Rees et al. 1982;
Wilson \& Colbert 1995; Sikora, Stawarz \& Lasota 2007), and may tell
us about the merger history of galaxies (e.g. Volonteri, Sikora \&
Lasota 2007).

In recent years it has become clear that many aspects of black hole
accretion and jet formation are directly comparable between AGN and
lower-mass (typically $\sim 10M_{\odot}$) black holes in X-ray binary
systems (XRBs). This is to be expected, given the very simple scalings
with mass for black holes in general relativity, although there is
likely to be a larger diversity of environments in AGN. Scalings
between mass, radio luminosity and X-ray luminosity are reported in
Merloni, Heinz \& di Matteo (2003) and Falcke, K\"ording \& Markoff
(2004; see also Koerding, Falcke \& Corbel 2006); scaling of fast
variability properties with mass and accretion rate are reported in
McHardy et al. (2006) and K\"ording et al. (2007); more qualitative
similarities between XRB and AGN accretion are noted in K\"ording,
Jester \& Fender (2006) and also discussed in Marscher et al. (2002)
and Chatterjee et al. (2009).  The temporal evolution of XRB jets,
relatively rapid compared to AGN, has allowed many estimates of the
power (e.g. Fender 2001; Gallo et al. 2005; K\"ording, Fender \&
Migliari 2006) and speed (e.g. Mirabel \& Rodriguez 1994;
Miller-Jones, Fender \& Nakar 2006) of the jets and their connection
to accretion `state' as characterized by the X-ray emission
(e.g. Fender et al. 1999; Fender, Belloni \& Gallo 2004 [hereafter
FBG04]; Corbel et al. 2004). Importantly these studies have shown that
the jet power of a black hole XRB, as well as the radiative
efficiency of the accretion flow, can change dramatically in the same
source at the same overall radiative luminosity on timescales far
shorter than those associated with significantly changing mass or
angular momentum.

In very brief summary, in black hole XRBs the coupling of radio
emission (and hence jets) to X-ray state and luminosity is as follows:
at Eddington ratios (in terms of X-ray luminosity) below about 0.01,
sources seem to be exclusively in the `hard' X-ray state in which the
X-ray emission is dominated by a component extending to $\sim 100$
keV, widely (but not universally) accepted to arise via thermal
Comptonisation of seed photons by a hot flow / corona. In this state
there is strong aperiodic variability and a steady, powerful,
flat-spectrum jet. The luminosities of the two components scales
roughly as $L_{\rm radio} \propto L_{X}^{0.6-0.7}$. At higher
Eddington ratios, reached generally by transient outbursting systems,
sources can switch into `softer' states in which the X-ray spectrum is
dominated by a cooler ($\sim 1$ keV) component with a near-blackbody
spectrum, generally interpreted as the inner accretion disc. In this
state the radio emission is either dramatically suppressed by a factor
$\geq 50$ or evolves to a fading, optically thin, state, both
scenarions suggesting the `quenching' of the core jet (possibly with
some remnant extended emission). In transitions from hard to soft
states major radio flares, often resolved as discrete, powerful,
ejection events, are commonly observed. Sources generally fade in the
soft state until they are once again at a few \% Eddington (in
$L_{X}$), and then make a transition back to the hard state in which
mode they fade further. The initial hard $\rightarrow$ soft state
transition is usually at a higher luminosity than the soft
$\rightarrow$ hard return branch, i.e. hysteresis when spectral
hardness is compared to luminosity.  Note that the same source has
been observed to make both hard $\rightarrow$ soft and soft
$\rightarrow$ hard transitions at different luminosities in different
outbursts; note further that some sources e.g. Cyg X-1 never drop
below the 1\% Eddington threshold and remain 'persistent and
variable'. For comprehensive reviews on these phenomena, see FBG04;
Remillard \& McClintock 2006; Done, Kubota \& Gierlinski 2007; Fender,
Homan \& Belloni 2009; Belloni 2009. The most comprehensive
compilation of X-ray data on black hole binaries is presented in Dunn
et al. (2010). Note that X-ray binary systems with comparable
properties but hosting a neutron star instead of a black hole
(candidate) also show jets, but with a lower ratio of $L_{\rm radio}$
to $L_{X}$ (Fender \& Kuulkers 2001; Migliari \& Fender 2006).

\begin{table*}
\begin{tabular}{ccccc}
\hline
Source & Mass & \multicolumn{2}{c}{Spin estimate} & Refs \\
       & $(M_{\odot}$ & Disc & Reflection &  \\
\hline
M33 X-7 & $15.6 \pm 1.5$ & $0.77 \pm 0.05$& & 1,6,7,17 \\
LMC X-1 & $10.9 \pm 1.4$ & $0.90^{+0.04}_{-0.09}$ & & 1,7,18 \\
LMC X-3 & $11.6 \pm 2.1$ & $<0.8$ & & 4,7, 19 \\
        &                & $-0.03$ & & 13 \\
GS 2000+25 & $7.2 \pm 1.7$ & 0.03 & & 1,13 \\
GS 1124-68 & $6.0 \pm 1.5$ &-0.04 & & 1,13 \\
4U 1543-47 & $9.4 \pm 1.0$ & 0.7--0.85 & $0.3 \pm 0.1$ & 1,2,3,7,8 \\
GRO J1655-40 & $6.30 \pm 0.27$ & 0.65--0.8 & $0.98 \pm 0.01$ & 1,2,3,7,9\\
             & & 0.93 & & 13 \\
GRS 1915+105 & $14 \pm 4$ & 0.98--1.0 & & 1,2,5,7 \\
             &            & 0--0.15 & & 10 \\
             &            & $\sim 0.7$ & & 11 \\
             &            & 0.998 & & 13 \\
XTE J1550-564 & 9.7--11.6 & $<0.8$ & $0.76 \pm 0.01$ & 1,4,7\\
XTE J1650-500 & $5 \pm 2$ & & $0.79 \pm 0.01$ & 1,7 \\
GX 339-4 & $\geq 6$ & &  $0.94 \pm 0.02$ & 1,7 \\
SAX J1711.6-3808 & & & $0.6^{+0.2}_{-0.4}$ & 7 \\
XTE J1908+094 & & & $0.75 \pm 0.09$ & 7 \\
Cygnus X-1 & $10 \pm 5$ & & $0.05 \pm 0.01$ & 1,7 \\
4U 1957+11 & 3--16 & 0.8--1.0 & & 1,12 \\
A 0620-00 & $6.6 \pm 0.3$ & $0.12^{+0.18}_{-0.20}$ & & 21 \\
\hline
MCG 6-30-15 & $(4.5 \pm 2) \times 10^6$  & & $0.989^+{0.009}_{-0.002}$ & 14 \\
SWIFT J2127.4+5654 & $\sim 10^7$ & & $0.6 \pm 0.2$  & 15 \\
Fairall 9 & $(2.6 \pm 0.6) \times 10^8$ & & $0.60 \pm 0.07$ & 16 \\ 
1H 0707-495 & $\sim 10^7$ & & $\geq 0.98$ & 20 \\
\hline
\end{tabular}
\caption{A compilation of published spin (and mass) measurements for
  black holes in both X-ray binary systems and AGN, based on disc and
  reflection/line measurements. All of these measurements, except
  those of Zhang et al. (1997; see text for discussion) and the two
  upper limits, are presented in Fig 1.  Ref 1 = Remillard \&
  McClintock (2006) and McClintock \& Remillard (2009), Ref 2 =
  McClintock, Narayan \& Shafee (2007), Ref 3 = Shafee et al. (2006),
  Ref 4 = Davis, Done \& Blaes (2006), Ref 5 = McClintock et
  al. (2006), Ref 6 = Liu et al. (2008), Ref 7 = Miller et al. (2009),
  Ref 8 = Gallo, Fender \& Pooley (2003), Ref 9 = fender, homan \&
  belloni and references therein, Ref 10 = Kato (2004), Ref 11 =
  Middleton et al. (2006), Ref 12 = Nowak et al. (2008), Ref 13 =
  Zhang et al. (1997), Ref 14 = Brenneman \& Reynolds (2006), Ref 15 =
  Miniutti et al. (2009), Ref 16 = Schmoll et al. (2009, Ref 17 =
  Orosz et al. (2007), Ref 18 = Orosz et al. (2009), Ref 19 =
  Val-Baker, Norton \& Negueruela (2007), Ref 20 = Fabian et
  al. (2009), Ref 21 = Gou et al. 2010 and references therein.}
\label{spintable}
\end{table*}

\begin{figure*}
\centerline{\epsfig{file=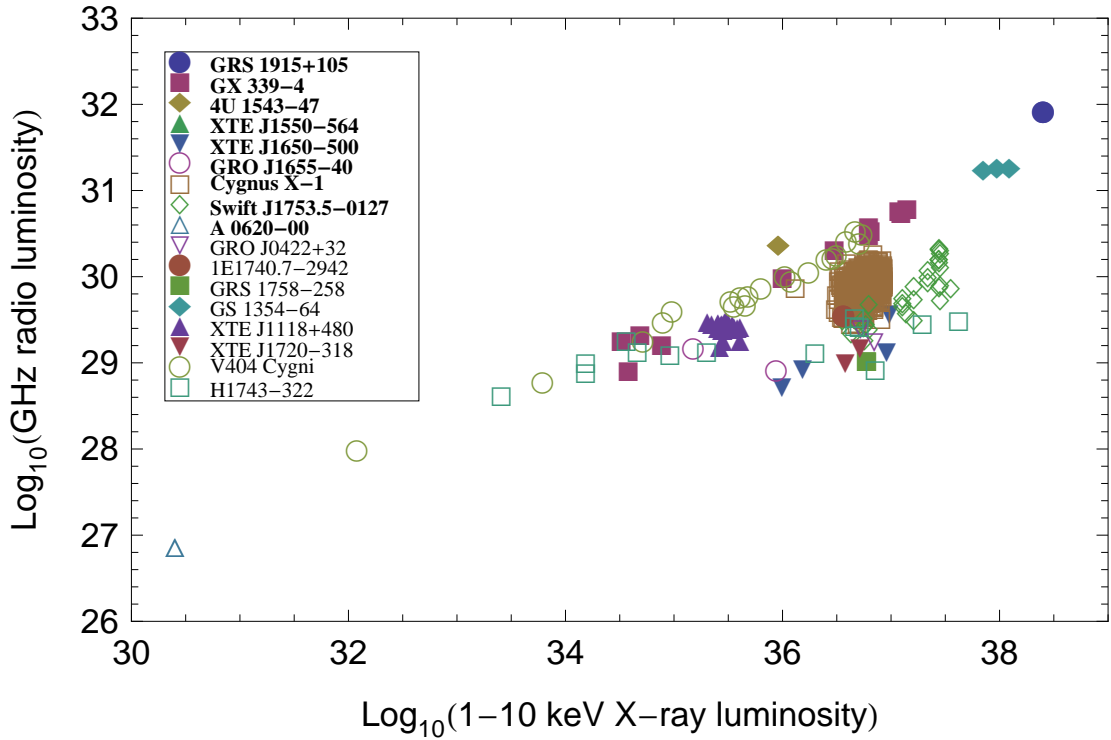, width=17cm, angle=0}}
\caption{The radio:X-ray plane for low/hard state black hole X-ray
  binaries. All currently available data are plotted, illustrating
  both the overall correlation over more than $10^8$ in X-ray
  luminosity, and also the increasing number of 'radio quiet' sources
  being found at relatively high X-ray luminosities. The first nine
  sources in the key, indicated in bold, have reported spin
  measurements (see Table 1). For those sources we have fitted a
  function with same slope as the ensemble (+0.6) but with variable
  normalization. In turn, we have used this normalisation as a measure
  of the relative jet power of the source, and compare it later to the
  reported spin measurements.}
\label{radio}
\end{figure*}

\begin{figure*}
\centerline{\epsfig{file=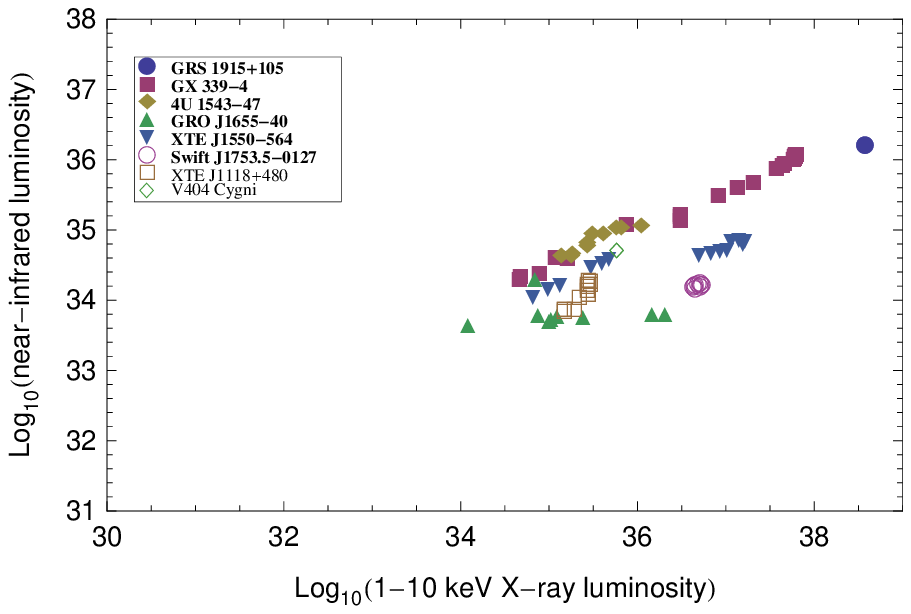, width=17cm, angle=0}}
\caption{As Fig\ref{radio} but for near-IR data. See Russell et
  al. (2006), Homan et al. (2005); Migliari et al. (2007); Russell et
  al. (2007b); for observational details. Note that there appear to be
  two tracks for the same source, XTE J1550-564 (solid blue inverted
  triangles), corresponding to the rise and decay phases of an
  outburst. This is an example of changing jet efficiency in the same
  source, same luminosity, same state, which is clearly not a spin
  effect. The two tracks for XTE J1550-564 are fitted separately in
  our analysis.}
\label{IR}
\end{figure*}

In parallel with these advances in the study of black hole jet power,
speed and relation to accretion state, there has been a rapid recent
growth in the number of estimates of spin of black holes in XRB
systems (the spin is generally discussed in terms of the dimensionless
spin parameter $a_* = cJ / GM^2$ which has a range from 0
[non-rotating, or `Schwarzschild' black hole] to 1 [maximally
  rotating, or `Extreme Kerr' black hole]). Two approaches have been
taken based on detailed fitting of X-ray spectra. In the first
approach, the accretion disc continuum is modelled; in the second
approach the `reflection' component, including the iron line around
6.4 keV, is also modelled. One of the earliest attempts to measure
spin from accretion disc continua was made by Zhang, Cui \& Chen
(1997), who reported that the two superluminal jet sources, GRS
1915+105 and GRO J1655-40 had high ($a_* > 0.9$) spin whereas three
other X-ray transients had much lower ($|a_*| < 0.05$), and possibly
in some cases retrograde (compared to the inner accretion disc)
spins. More recent disc-modelling results have been presented by,
amongst others, Shafee et al. (2006), McClintock et al. (2006), Davis,
Done \& Blaes (2006), Middleton et al. (2006), Nowak et al. (2008),
Steiner et al. (2009), Gou et al. (2010). Recent results from
modelling of the reflection component are compiled in Miller et
al. (2009). General points to take from the presented results are
reports of very high spins for some black holes (e.g. 0.98--1.00 for
GRS 1915+105 from disc measurements in McClintock et al. 2006; $0.98
\pm 0.01$ for GRO J1655-40 from reflection components in Miller et
al. 2009), some discrepancies (see discussion in Miller et al. 2009
and our table 1), and low spin measurements for both Cygnus X-1 ($0.05
\pm 0.01$; Miller et al. 2009) and A 0620-00 ($0.12^{+0.18}_{-0.20}$;
Gou et al. 2010).  Several criticisms of the spin-fitting methods have
appeared in the literature (e.g. Kolehmainen \& Done 2010; Done \&
Diaz-Trigo 2010). In the context of disc-fitting, we further note that
Fragile (2010) has reported that fits to a system where the black hole
spin and inner accretion disc axes are misaligned by only 15$^{\circ}$
are enough to render essentially useless inferred measurements of spin
via this method.

\begin{table}
\begin{tabular}{ccc}
\hline
Source & Distance (kpc) & Refs \\
\hline
GRS 1915+105 & 11.0 & 1 \\
GX 339-4 & 8.0 & 2 \\       
4U 1543-47 & 7.5 & 2 \\
XTE J1550-564 & 5.3 & 3 \\
XTE J1650-500 & 2.6 & 4 \\
GRO J1655-40 & 3.2 & 2 \\
Cygnus X-1 & 2.1 & 1 \\
Swift J1753.5-0127 & 8.0 & 4\\
GRO J0422+32 & 2.5 & 2 \\
1E1740.7-2942 & 8.5 & 1 \\
A 0620-00 & 1.2 & 2 \\
GRS 1758-258 & 8.5 & 1 \\
GS 1354-64 & 25.0 & 5\\
XTE J1118+480 & 1.7 & 2 \\
XTE J1720-318 & 6.5 & 6 \\
V404 Cyg & 2.4 & 7\\
H1743-322 & 7.5 & 8\\
\hline
\end{tabular}
\caption{Source distances adopted in this paper. Ref 1 = Gallo, Fender
  \& Pooley (2003) and references therein; Ref 2 = Russell et
  al. (2006) and references therein; Ref 3 = Hannikanen et al. (2009);
  Ref 3 = Homan et al. (2006); Ref 4 = Zurita et al. (2008); Ref 5 =
  Casares et al. (2009); Ref 6 = Chaty \& Bessolaz (2006); Ref 7 =
  Miller-Jones et al. (2009); Jonker et al. (2010).}
\end{table}

In this paper we take these reported measurements of black hole spin
and compare them against different methods of estimating the power
and, in some cases, speed of the jet observed in such systems. From
these comparisons we will draw conclusions about evidence for the
dependence of jet power, or speed, on spin, in accreting black holes.
Note that in this paper we are not considering estimates of black hole
spin based upon other methods, such as frequencies of quasi-periodic
oscillations.

\section{X-ray binaries}

\begin{figure*}
\centerline{\epsfig{file=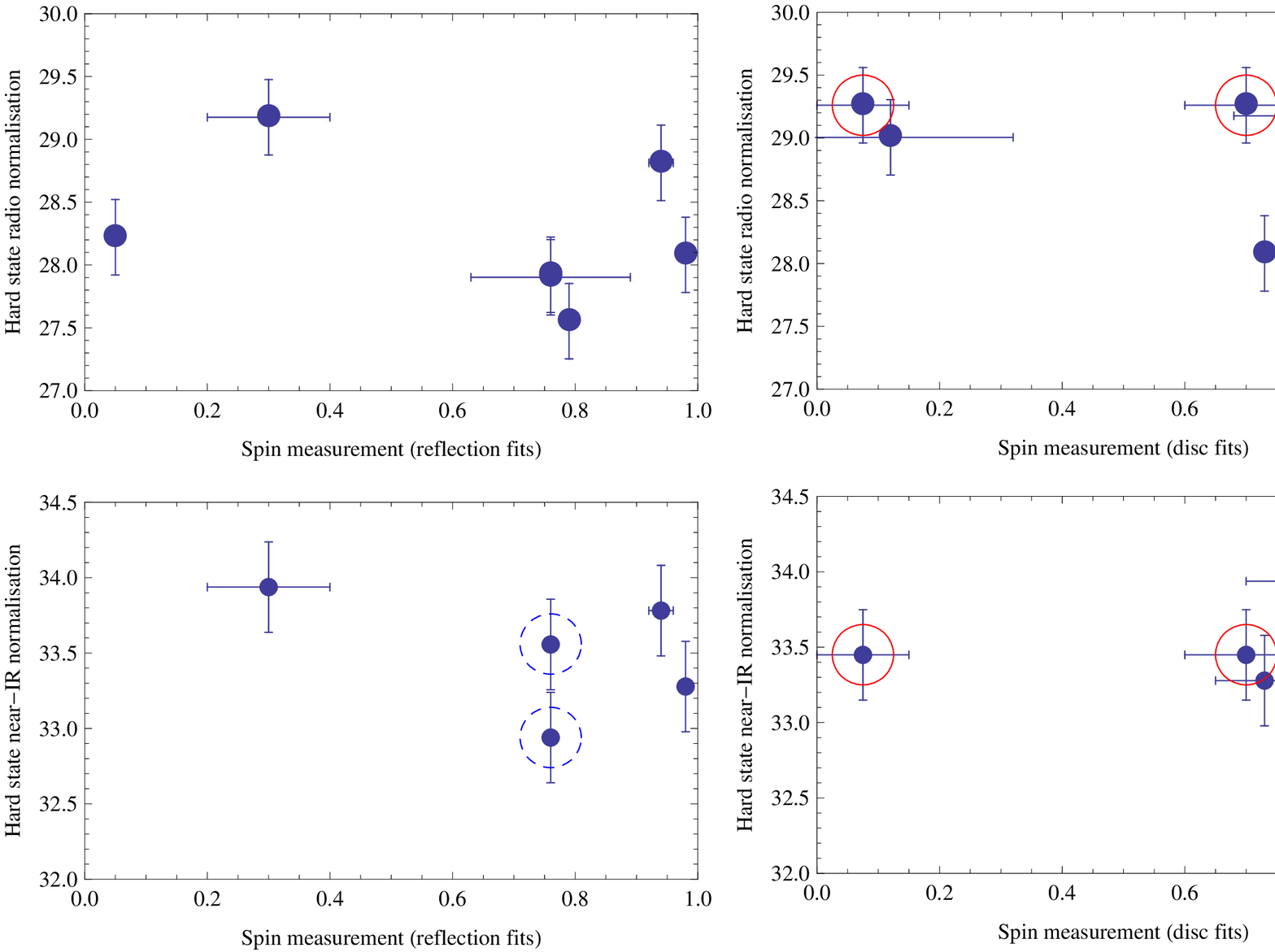, width=18cm, angle=0}}
\caption{A comparison of the jet power normalisations found from radio
  (upper) and near-infrared (lower) with reported black hole spin
  measurements, from reflection (left) and disc fits (right). Despite
  reportedly sampling the entire range of black hole spins there is
  clearly no dependence of jet power on these reported values. The
  left-oriented arrows in the disc fits indicate the upper limit of
  $\leq 0.8$ reported for the spin of XTE J1550-564 based on disc
  measurements. Note that in the near-IR jet power panels, XTE
  J1550-564 has two measurements, based on the different apparent jet
  power normalisations in the rise and decay phases of an outburst;
  these are indicated by dotted circles (and are included to
  demonstrate the range of currently inexplicable apparent changes in
  jet production efficiency). GRS 1915+105 has three reported spin
  measurements, which are all plotted, indicated by solid red
  circles.}
\label{spincomp}
\end{figure*}

Table \ref{spintable} lists all the reported measurements of black
hole spin which we have been able to find for X-ray binary systems, as
well as the four reported spin measurements for AGN (not counting
spins inferred for entire populations of AGN based on distributions of
radio loudness and/or radiative efficiency). As noted above there are
some intriguing claims, notably that the X-ray binaries GRS 1915+105
and GRO J1655-40, as well as the AGN MCG 6-30-15 and 1H 0707-495, have
very high spins, whereas the black hole binaries Cyg X-1 and A 0620-00
have a very low spins.  Figure 1 summarizes using histograms the
current distributions of reported spins; clearly there is a bias
towards higher spin measurements, which is to be expected since these
cases should correspond to the strongest observational effects.  In
the following we shall compare these reported spin measurements with
estimates of the jet power in the hard state, and both jet speed and
power in transient outbursts. Most of the sources with reported spin
measurements have radio and/or near-infrared measurements which allow
estimates of the jet power. Note that in these histograms and the
subsequent analyses, we do not use the spin measurements reported by
Zhang et al. (1997), although we do list them in table 1. This is
because they are likely to have been superceded by more recent
refinements of the disc-fitting method, although in some cases their
measurements are in agreement with more recent fits (see
e.g. discussion in McClintock et al. 2006).

\subsection{The hard state jet}

In the hard state, we can only really compare jet power, and not
speed, between sources to see if it correlates with estimates of black
hole spin. We note that the analyses of Gallo, Fender \& Pooley (2003)
and Heinz \& Merloni (2004) already indicate that the range of Lorentz
factors of such hard state jets is likely to be small (although the
absolute value is as yet undetermined). Therefore we can immediately
conclude that if the reported range of spins in the hard state,
0.05--0.99, is correct then the speed of jets in the hard state does
not have a strong dependence on black hole spin (unless the jets are
sub-relativistic, when a dependence on spin would not be measurable in
terms of Doppler boosting effects).

A variety of approaches may be taken to estimate the power of BH XRB
jets in the hard X-ray state (e.g. Fender 2001; Fender, Gallo \&
Jonker 2003; Malzac, Merloni \& Fabian 2004; Gallo et al. 2005;
K\"ording, Fender \& Migliari 2006); nearly all result in high
normalisations for the relation, such that in bright (typically in the
range $10^{-3}$ to $10^{-1}$ Eddington in X-ray luminosity) hard
states the radiative (X-ray) and kinetic luminosities are
comparable. At lower luminosities in the hard state the jet probably
comes to dominate over the X-ray emission (Fender, Gallo \& Jonker;
see also discussion in Cabanac et al. 2009). In particular, K\"ording,
Fender \& Migliari (2006) present a summary of jet power as a function
of accretion rate for a small sample of both black hole and neutron
star binaries. There may be a slightly higher rate of jet power per
unit accreted mass (a factor of order unity) for the black
holes. Importantly, both Cyg X-1 and GRS 1915+105 are in the sample
and there is no evidence that there is any difference in the
normalization of the jet power as a function of accretion rate between
them, to a level of within a factor of two.

However, we can test the relation between jet power and black hole
spin estimates for a sample of BH XRBs more explicitly.  Under the
assumption (reasonable, but not proven) that the radio through
infrared spectral energy distributions of BH XRBs in the hard state
are the broadly same from system to system (although a varying
function of accretion rate), we can use measurements of jet emission
in different bands to compare the relative power of jets between
sources. Note that the relation between the observed flat-spectrum
synchrotron luminosity and total jet power is believed to be of the
form $L_{\rm Radio} \propto L_{\rm Jet}^{1.4}$ (Blandford \& K\"onigl
1979; see e.g. K\"ording, Fender \& Migliari 2006 for observational
support for this scaling). Recent simulations of X-ray binary jets
with internal shocks appear to reproduce this scaling (Jamil, Fender
\& Kaiser 2010).

In Fig \ref{radio} we present the most up to date compilation of
quasi-simultaneous radio and X-ray observations of hard state black
hole XRBs; table 2 lists the distances adopted for this plot -- some
are notably different from those used in Gallo, Fender \& Pooley
(2003). Note that (as with the near-infrared, see below) the radio flux
densities have been multiplied by the frequency of the radio
observations (typically 5--8 GHz) to give an estimate of the radio
power. It is clear that the `universal' correlation reported in Gallo,
Fender \& Pooley (2003) is in fact something much broader, at least at
high luminosities. In fact without Cyg X-1, the plot looks rather like
there are two distinct tracks, reminiscent of the `radio loud':`radio
quiet' divide in AGN. In order to make a uniform estimate of the jet
power for each of these sources, we fit a straight line to each
system; the slope of the fit is fixed to +0.6 (i.e. $L_{\rm radio}
\propto L_{\rm X}^{0.6}$) as found in Gallo et al. (2006) for the most
recent ensemble analysis, so we are simply fitting normalisations to
the relation as a proxy for relative jet power. Note that for 4U
1543-47, XTE J1550-564 and A 0620-00 we only have a single datum, and so the
`fit' is a simple scaling, no more. Note also that we do not consider
upper limits, and that for Cygnus X-1 we do not include points which
include any evidence for suppression of the radio emission as the
source enters softer X-ray states (see Fig 3 of Gallo, Fender \&
Pooley 2003 for an illustration of this). The normalisations, $c$, are
simply fitted as

\[
\log_{10} L_{\rm radio} = c + 0.6 (\log_{10} L_{\rm X} - 34)
\]

This process can be repeated with near-infrared data, which have been
convincingly demonstrated to have a large contribution from the jet
(e.g. Homan et al. 2005; Russell et al. 2006). In Fig \ref{IR} we plot
the equivalant ensemble of near-infrared data, and perform the same
analysis of normalisations. For XTE J1550-564 we plot data both in the
rise and decline phases of an outburst, which show different
normalisations -- see Russell et al. 2007 and our discussion
later. Note also that the correlation in Russell et al. (2006) extends
to lower luminosities because it also utilizes optical data; however
those data are generally dominated by the irradiated accretion disc
and are not suitable for estimating the jet power.

For both the radio and infrared data sets, we include a
`representative' measurement for the hard `plateau' state of GRS
1915+105 (Fender \& Belloni 2004). These measurements should be
interpreted with caution as this system -- persistently very luminous
since entering outburst in 1992 -- has not been observed to enter
a true canonical hard state. Nevertheless, the properties of the
source in this plateau state (which is probably a `hard intermediate'
state in the terminology of Belloni 2009), including a steady powerful
radio jet, are rather similar to those of the canonical hard state.

We can now compare these measurements of the radio and near-IR
normalisations, as proxies for jet power, with the reported
measurements of black hole spin from reflection and disc modelling.
This is done in Fig \ref{spincomp}, where for each normalisation
measurement we estimate a systematic uncertainty of 0.3 dex.  There is
clearly no correlation in any of the four panels. Notably, for the
reflection fits, Cyg X-1 appears to have more or less average radio
power despite a low reported spin. Equally, A 0620-00 has a strong
radio normalisation (admittedly based on a single measurement),
compared to a low reported spin from disc fits.  Note that we indicate
(with solid red circles) all three of the other reported spin
measurements for GRS 1915+105.  The lower panels also clearly
illustrate the large difference in relative jet power fitted to the
source XTE J1550-564 (indicated with dashed blue circles) when fitting
either the rise (lower measurement) or decay (upper measurement)
phases of an outburst.

It is important to note that while there are considerable
uncertainties in the absolute normalisation and form of the relation
betwen radio luminosity and total jet power what we have measured here
is a fairly well-defined {\em ranking}. In this context it is
important to note that the source with the lowest reported spin,
Cygnus X-1, is also one of the best constrained, being at a relatively
small distance and with detailed studies of the jets (Gallo et
al. 2005; Heinz 2006). Note also that XTE J1650-500 (see Corbel et
al. 2004 for more details) clearly shows the pattern of the global
correlation, but at a lower normalisation than the rest (Fig 2),
despite having a relatively high reported spin ($0.79 \pm 0.01$).

Overall, we conclude that while there may be evidence for the
requirement of an additional parameter determining jet power in hard
state black hole binaries, such a parameter in no way correlates with
reported estimates of black hole spin. It is worth noting that while
there may be some unknown systematics, which may exceed our 0.3 dex
estimate it seems very unlikely indeed that these systematics could be
enough to hide a genuinely strong trend with reported spin.

\subsection{Powerful, transient jets}

For the powerful, transient, jets we may potentially explore both jet
power and jet speed (since we have proper motions in several cases) as
functions of estimated black hole spin (whereas for the hard state
jets there are no clear speed measurements), although as we will see
below we only really have lower limits to the jet speeds and cannot
make much progress.

\subsubsection{Transient jet power}

It is not straightforward to measure the power associated with the
transient ejection events. Typically we calculate the minimum energy
associated with some synchrotron event, and divide by the rise time to
get the average power going into the jet. This approach is useful to
provide lower limits on, and order of magnitude estimates of, jet
power but is very susceptible to errors resulting from poor sampling
of events, uncertainties in Doppler boosting, assumptions about
equipartition etc. As a result both the normalisation and ranking of
jet powers between different sources is less accurate than for the
hard state. Nevertheless we can make a comparison, and for this
purpose we will use the transient jet powers estimated in FBG04,
compared with the spin measurements compiled in this paper.

\begin{figure}
\centerline{\epsfig{file=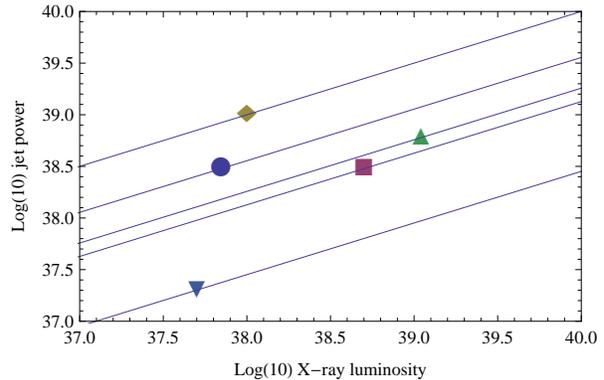, width=8cm, angle=0}}
\caption{Comparison of estimated radio jet power as a function of
  X-ray luminosity, using data from FBG04. Each point is for a
  different source; the fitted lines correspond to a fixed slope of
  +0.5 as fitted for a broader ensemble in FBG04.}
\label{transpower}
\end{figure}

In Fig {\ref{transpower}} we plot as a function of X-ray luminosity the
estimated transient jet powers for five systems listed in FBG04 for
which there are reported spin measurements. The fitted lines are of
fixed slope +0.5 (as fitted to the ensemble of transient jet powers by
FBG04), and so we may compare the fitted normalisations in a process
analogous to that employed for the hard state radio and near-IR
measurements earlier in the paper. The normalisations, $c$, in this
case are:

\[
\log_{10} L_{\rm jet} = c + 0.5 (\log_{10} L_{\rm X} - 34)
\]

\begin{figure*}
\centerline{\epsfig{file=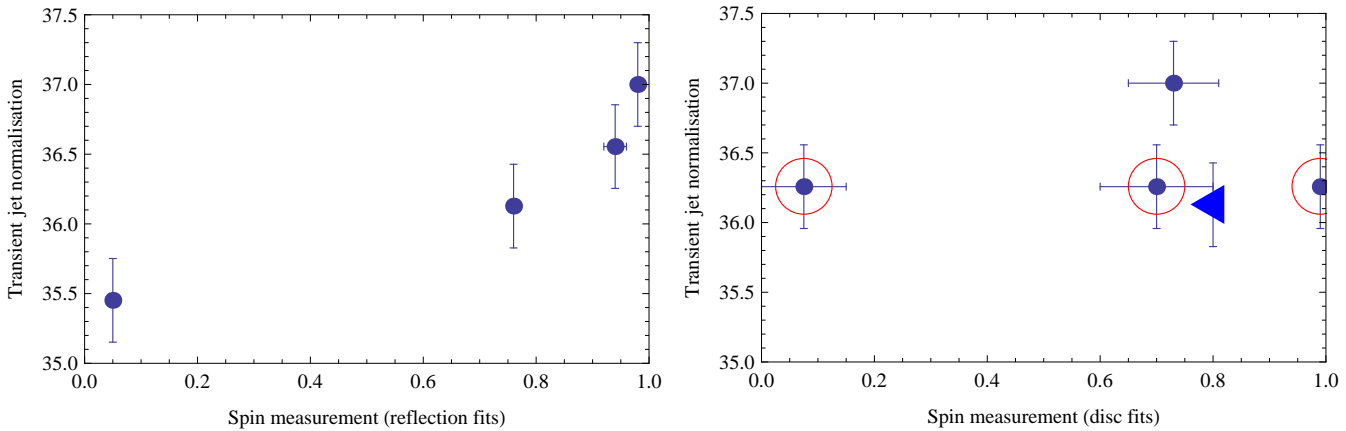, width=18cm, angle=0}}
\caption{Comparison of transient jet power normalisations with
  reported spin measurements from reflection (left) and disc
  (right) fits.}
\label{transpowerspin}
\end{figure*}

In Fig {\ref{transpowerspin}} we compare these fitted normalisations
with the reported measurements of black hole spin. While the disc
measurements again show no correlation with the estimated jet power,
there is an intriguing apparent correlation between jet power and spin
for the reflection measurements. We caution the reader not to
over-interpret this, given all the uncertainties outlined above, and
discuss it further in section \ref{discussion}.

\subsubsection{Transient jet speed}

As discussed already in Fender (2003) and FBG04, in nearly all cases
it is only possible to place a lower limit on the speed of jets from
X-ray binaries when basing the estimates on measurements of proper
motions alone. This is because the distance uncertainties typically
encompass a range of possible solutions for the Lorentz factor from $2
\leq \Gamma \leq \infty$.  A different approach was taken in
Miller-Jones, Fender \& Nakar (2005) in which estimates of the jet
Lorentz factor were made under the assumption of free relativistic
expansion in the rest frame of the jet, with time dilation causing the
apparently very small opening angles (i.e. retarded apparent
expansion).  A third approach to estimating the Lorentz factor of jets
is available from the ratio of approaching to receding
jets. Unfortunately in all cases we still end up with lower limits on
the Lorentz factor.

All of these three approaches have their uncertainties; we summarize
these estimates, for sources with reported spin measurements, in table
\ref{speed}. Once again, there is no evidence for any correlation with
reported black hole spin measurements. Furthermore it is worth noting
that currently the highest speed measured for a jet from an X-ray
binary is that from the {\em neutron star} system Circinus X-1 (Fender
et al. 2004; Tudose et al. 2009), although this may be a jet quite
unlike those observed from accreting black holes.

\begin{table}
\begin{tabular}{cccc}
Source & $\Gamma_{\mu}$ & $\Gamma_{exp}$ & $\Gamma_{ratio}$ \\
\hline
GRS 1915+105 (1997) & $\geq 2$ & $\geq 15.7$ & $\geq 1.8$ \\
GRS 1915+105 (2001) & & $\geq 3.3$ & \\
GRS 1915+105 (steady) & & $\geq 11.7$ & \\
GRO J1655-40 & $\geq 1.7$ & $\geq 18.6$ & $\geq 2.4$ \\
XTE J1550-564 & $\geq 2$ & $\geq 19.6$ & $1.3 \pm 0.2$ \\
Cyg X-1 & $\geq 1.3$ & $\geq 33$ & \\
GX 339-4 & $\geq 2.3$ & $\geq 4.9$ & \\
\hline
\end{tabular}
\caption{Estimates of jet Lorentz factor ($(1-\beta^2)^{-1/2}$, where
$\beta=v/c$) for black hole X-ray binaries with reported spin
measurements, from three different methods (see main text).}
\label{speed}
\end{table} 

Because these measurements are all lower limits, it is more or less
impossible to attempt any correlation with the reported spin
measurements. Nevertheless, it is worth bearing in mind that there is
evidence that transient jets are faster than the hard state jets
(FBG04) and so it may be that this boost in speed is somehow connected
to the black hole spin (but in this case the jet in Cyg X-1 should be
slower than those from other transients).

\section{Discussion}
\label{discussion}

We have clearly demonstrated that for the black hole X-ray binaries in
the hard X-ray state there is no correlation between the reported spin
measurements and either jet power or speed. If the spin measurements
are correct then any dependence on spin which does exist in the hard
state must be very weak, less than about an order of magnitude across
the whole range of spins from 0.05 -- 0.98.  Recall that in McKinney
(2005) a jet with $a_* \geq 0.9$ should have a jet efficiency more
than $10^4$ greater than one with $a_* \leq 0.2$ (although also recall
that Livio, Ogilvie \& Pringle [1999] argue the opposite position,
that the spin cannot be efficiently tapped via this process). Note
that McKinney's $a_*^4$ dependence of jet power on spin is much
steeper than the $a_*^2$ originally estimated by Blandford \& Znajek
(1977).

\subsection{Different types of jets -- only one spin-powered?}

It has been suggested (e.g. Meier 1999; FBG04; see also
Tchekovskoy et al. 2010) that the two apparently different types of
jets in BH XRBs (slow and steady in hard state, fast and transient at
hard $\rightarrow$ soft state transitions; both very powerful) may be
powered in different ways. In particular it has been suggested that
the hard state jet may be powered by the disc (via e.g. the
centrifugal mechanism of Blandford \& Payne 1982), while only the
transient jets arise close enough to the black hole to be affected by
spin, and possibly the Blandford-Znajek (or related) process.  This
suggestion is interesting in the context of AGN, because lower
Eddington ratio systems seem in fact to be responsible for most of the
kinetic feedback in the universe (e.g. Merloni, Heinz \& Schwab 2007;
Koerding, Jester \& Fender 2007). In other words, even if the
transient jets are in some way spin-powered, kinetic feedback in the
universe over cosmological time has not been strongly influenced by
this spin. Furthermore, it is the lower Eddington-ratio AGN which are
the sources which Sikora et al. (2007) claim to show the clearest
evidence for a radio loudness bimodality (which they attribute to
spin), seemingly contrary to our results for the hard state.

With these caveats in mind, we can still explore if there might be a
transition to spin-powered jets in transient states, as possibly
suggested by the left panel of Fig 6, with the contribution from spin
(as measured via reflection) increasing the jet power by about an
order of magnitude for the extreme Kerr black holes. Such an
interpretation would not be arrived at, obviously, if the disc fits
were also included. If this hypothesis were the case, then we would
expect a step up in jet power for the transient systems at the point
at which they `connect' to the spin. Whether or not such a step-up in
jet power between hard states and transient jets existed was explored
in FBG04 (their Fig 5), which shows that if the lower limit to jet
power of Fender et al. (2003) is correct, then there may be a boost of
about an order of magnitude in jet power for the transient
events. However, in the same figure the jet power normalisation from
Malzac et al. (2004) is also plotted, which does not require any
step-up in jet power. Since then, the jet power estimates for Cyg X-1,
based upon the apparently jet-blown cavity in the ISM, are very close
to the normalisation of Malzac et al. (Gallo et al. 2005; Heinz 2006;
Russell et al. 2007a). Surveying all of the available evidence we
conclude that there is no strong evidence for a dependence of jet
power on the reported black hole spin although there is room for a
weak dependence in the case that only the reflection measurements are
correct.

\begin{table*}
\begin{tabular}{|c|c|c|c|}
\hline
\multicolumn{2}{c}{Summary of evidence for reported black hole spin influencing jet} \\
\hline
 & {\bf Hard state jet} & {\bf Transient jet} & {\bf Soft state (suppressed) jet} \\
\hline
{\bf Jet Power} & Strong evidence against & Moderate evidence against & Weak evidence against \\
                & (from radio:X-ray correlations) & (from jet power:X-ray correlations) &  (from radio:X-ray correlations and AGN) \\
\hline
{\bf Jet Speeds} & Strong evidence against & Weak evidence against & No evidence \\
                 & (Narrowness of radio:X-ray sample distribution) & (but only lower limits to speed) & \\
\hline
\end{tabular}
\caption{A summary of our conclusions on any relation between reported
  black hole spin measurements (mainly for X-ray binaries) and the
  power and speed of observed jets. There is no good evidence for a
  connection in any aspect, although some are poorly tested.}
\label{summ}
\end{table*}

\subsection{Changing jet power without changing spin}

In black hole X-ray binaries there are two well-established circumstances in
which the jet power can change significantly in the same source on
short timescales, i.e. without any possible significant change in the
black hole spin. The first of these is well-established for a decade
now, and was in fact observed during the very first black hole state
change, that of Cygnus X-1 in 1972 (Tanabaum et al. 1972). It is the
observation that the radio emission in many black hole systems drops
dramatically in the {\em soft} X-ray state, compared to the same
levels of X-ray emission in the {\em hard} X-ray state (Fender et
al. 1999; Fender, Belloni \& Gallo 2004; Corbel et al. 2004). This
decrease in radio emission has been observed to be by at least a
factor of 50. It is important to note
that there are in fact many detections of the radio emission in soft
X-ray states, but in the majority of cases the emission is optically
thin and fading, consistent with (but not proof of) an origin in
previously-ejected material and not a core radio jet (see e.g. Corbel
et al. 2002; Fender, Homan \& Belloni 2009). In addition to the radio
measurements, the decrease in near-infrared jet emission at the
transition to the soft X-ray state is also clearly observed, in some
cases by up to a factor of 20 (Homan et al. 2005; Russell et
al. 2006; see also Russell \& Fender 2010).

In addition to this, there are two less well known effects which have
only more recently become clear. The first is that even in hard X-ray
states, in the same BH XRB, there can apparently be a range of jet
powers at the same X-ray luminosity. This is well-observed in the
near-infrared jet emsission of XTE J1550-564 (Russell et al. 2007),
where the near-infrared emission from the jet returns at a {\em
  higher} level post-outburst, at the same X-ray luminosity in the
same state, by a factor $\sim 5$ (Russell et al. 2007; this effect is
visible in our Fig 3). The same physical effect, this time of the same
source making 'parallel tracks' in the hard state radio vs. X-ray
correlation, is also in the case of GX 339-4 (Corbel et al. in prep;
see also Coriat et al. 2009).

Thirdly, it has been shown in recent years that several black hole
XRBs seem to be a long way below the `universal' radio:X-ray
correlation reported for the hard state by Gallo, Fender \& Pooley
(2003). Gallo (2007) and Cadolle-Bel et al. (2007) present clear
examples of this, and it is apparent in our Fig 2. The origin of these
discrepancies, is unclear but could be related to errors in distance
estimates, combinations of orientation and beaming, or simply
efficiency of jet production (which may in turn be related to the
magnetisation of the jet -- see Casella \& Pe'er 2009). Interestingly,
two of the sources which lie significantly below the correlation,
i.e. are apparently 'radio quiet', have spin measurements reported by
Miller et al. (2009). XTE J1650-500 ($a_* = 0.79 \pm 0.01$) and 4U
1543-57 ($a_* = 0.3 \pm 0.1$) both lie significantly below the
correlation. Unfortunately these spins lie between those measured by
the same authors for GX 339-4 ($a_* = 0.94 \pm 0.02$) and Cyg X-1
($a_* = 0.05 \pm 0.01$), {\em both of which} have a higher
normalisation in the $L_{radio}$:$L_{\rm X}$ plane.

In summary, based on X-ray binaries, there is evidence for at least
one further parameter which affects the efficiency of jet and/or
radiation production in an accreting black hole system, but this
parameter is not associated with black hole spin.

\subsection{Relation to AGN}

The strongest case made in recent years for spin-powering of some AGN
jets is that put forward by Sikora et al. (2007), who demonstrate an
apparent bimodality in the relation of radio loudness as a function of
Eddington ratio for different classes of AGN (see also e.g. Lal \& Ho
2009). They find that while there is an overall trend of decreasing
radio loudness with increasing luminosity for all classes of AGN,
there appear to be two distinct, approximately parallel tracks, with
broad-line radio galaxies (BLRGs), radio-loud quasars, and FR~I objects
on the upper track, and PG quasars, Seyferts and LINERs on the lower
track. This effect is most evident at the lower Eddington ratios --
above $\sim 0.01$ Eddington some BLRGs are actually radio-quiet and
some PG quasars radio-loud. Sikora et al. attribute the parallel
tracks to representing populations of high-spin (radio loud) and
low-spin (radio-quiet) black holes respectively; a key point is that
all of the objects on the high-spin track reside in large
ellipticals. The partial mixing of radio loudnesses at the highest
Eddington ratios they attribute to a mixture of states, in analogy
with the behaviour of X-ray binaries (FBG04). Therefore in their
scenario, radio loudness above about 1\% Eddington results from a
combination of both spin and accretion state, whereas below this level
it is dominated by spin. There are reasonable arguments for the
evolution of black hole spin resulting from merger histories to
support such an interpretation (e.g. Volonteri, Sikora \& Lasota 2007
and references therein).

However, as noted in Sikora et al. (2007), the apparent bimodality in
radio loudness at low luminosities probably only arises when total and
not core radio luminosities are taken into account (see Terashima \&
Wilson 2003). The implication is therefore that somehow high-spin
systems produce the same core radio luminosity and only reveal their
more powerful jets by their stronger interaction with the ambient
medium (see e.g. discussion in Zirbel \& Baum 1995). This seems at
first counter-intuitive to us, based on our experience with X-ray
binaries, and would require that the relation between jet power and
radio emission (which seems so evident in the hard state of black hole
X-ray binaries) become `saturated' above some luminosity. The
`fundamental planes of black hole activity' presented by Merloni et
al. (2003) and Falcke et al. (2004) similarly only use core radio
luminosity, and do not find strong evidence for a radio loudness
bimodality.  Presumably other factors such as jet lifetime and the
properties of the surrounding medium must also play a role in the
strength of the extended emission (it is worth noting that the
strongest such `lobe' emission from a black hole X-ray binary is
probably that associated with the apparently low-spin black hole in
Cyg X-1).

\section{Conclusions}

So, is there any case for the reported black hole spins being
correlated with jet power or jet velocity in black hole X-ray binaries
? Almost certainly no. Our view on the evidence is summarised in Table
\ref{summ}.  This leads us to conclude that either:

\begin{enumerate}
\item{One or more of the methods used for estimating jet power or velocity are in error}
\item{One or more of the methods used for estimating black hole spin are in error}
\item{Jet power and/or velocity are {\em not} related to black hole spin}
\end{enumerate}

In addition to this lack of observational evidence for a relation
between black hole spin and jet power or speed, we have highlighted
the fact that there appear to be at least three different ways in
which the jet power and/or radiative efficiency -- both of which in
the context of AGN are used as estimators of spin -- from a black hole
X-ray binary may vary. Two of which are certainly independent of spin
because they occur in the same source on relatively short timescales,
and the third which does not correlate with any reported measurements
of black hole spin.

This paper is not setting out to argue that black hole spin does not,
in some cases, affect the power or speed of jets formed by that black
hole. However, current estimates of all three parameters (spin, jet
power, jet speed) of black hole X-ray binaries show no evidence for a
strong relation between them. Furthermore, it is suggested that as
well as pursuing the spin--jet connection, researchers working on AGN
populations should consider more carefully the fact that observations
of black hole binaries suggest there may be parameters other than spin
which determine the radio loudness of a system.

\section*{Acknowledgements}

We would like to thank Paolo Soleri for providing the full data set
for Swift J1753.5-0127.  We would also like to thank both referees of
the paper for their stimulating reports. RPF would like to thank Phil
Uttley, Luis Ho, Gilles Henri, Pierre-Olivier Petrucci, Jonathan
Ferreira and Guillaume Dubus for stimulating discussions.  RPF
acknowledges support from a Leverhulme Research Fellowship. EG is
supported through Hubble Postdoctoral Fellowship grant number
HST-HF-01218.01-A from the Space Telescope Science Institute, which is
operated by the AURA, Inc., under NASA contract NAS5-26555. DMR
acknowledges support from a Netherlands Organization for Scientific
Research (NWO) Veni Fellowship.

\end{document}